\documentclass{sigchi}

\usepackage{arabtex}
\usepackage{utf8}
\usepackage{balance}  
\usepackage{graphics} 
\usepackage{txfonts}
\usepackage{times}    
\usepackage[pdftex]{hyperref}
\usepackage{url}      
\usepackage{color}
\usepackage{textcomp}
\usepackage{booktabs}
\usepackage{ccicons}
\usepackage{todonotes}
\usepackage{relsize}
\usepackage{float}

\makeatletter
\def\url@leostyle{%
  \@ifundefined{selectfont}{\def\UrlFont{\sf}}{\def\UrlFont{\small\bf\ttfamily}}}
\makeatother
\def\pprw{8.5in}
\def\pprh{11in}

\definecolor{linkColor}{RGB}{6,125,233}
\hypersetup{%
  pdftitle={SIGCHI Conference Proceedings Format},
  pdfauthor={LaTeX},
  pdfkeywords={SIGCHI, proceedings, archival format},
  bookmarksnumbered,
  pdfstartview={FitH},
  colorlinks,
  citecolor=black,
  filecolor=black,
  linkcolor=black,
  urlcolor=linkColor,
  breaklinks=true,
}

\begin{document}
\sloppy
\setcode{utf8}
\title{Attitudes towards Refugees in Light of the Paris Attacks}

\numberofauthors{1}
\author{%
 \alignauthor{Kareem Darwish, Walid Magdy}\\
    \affaddr{Qatar Computing Research Institute, HBKU}\\
    \affaddr{Doha, Qatar}\\
    \email{\{kdarwish,wmagdy\}@qf.org.qa}\\
}

\maketitle

\begin{abstract}
The Paris attacks prompted a massive response on social media including Twitter. This paper explores the immediate response of English speakers on Twitter towards Middle Eastern refugees in Europe.  We  show that antagonism towards refugees is mostly coming from the United States and is mostly partisan.
\end{abstract}



\section{Introduction}
The coordinated terrorist attacks on multiple spots in Paris on Friday November 13, 2015 prompted a massive worldwide response on social media including Twitter.  The response was mostly focused on expressing outrage at the attacks and sympathy for the victims.  Other aspects of the response pertained to attitudes towards refugees, particular Syrian refugees.  These attitudes varied widely from those disassociating refugees from the perpetrated crimes to those who are blaming them for the crimes and calling to block their entry into Western countries.  In this paper, we explore a large set of 8.5 million tweets that we collected over 50 hours after the attacks. We collected tweets that contained words and hashtags that were relevant to the event such as \#pray4paris, \#parisAttacks, \#notInMyName, \#AttaqueParis, etc.

\section{Data Collection}
We continuously collected tweets between 5:26 AM (GMT) (roughly 7-8 hours after the attacks) on November 14 and 7:13 AM (GMT) on November 16 (approximately 50 hours in total) by searching the Twitter Rest API using a set of terms that are related to the terrorist attacks. These terms were: Paris, France, terrorist, terrorism, Muslims, Islam, Muslim, Islamic, terrorists, PorteOuverte, ParisAttacks, pray4Paris, paryers4Paris, AttaquesParis.  In total we collected 8.36 million tweets.  Since we were using the public APIs and due to Twitter limits, we were sampling the Twitter stream and our samples were reaching the set limits. In all, we were collecting 140k to 175k tweets per hour.  Also, since we were using mostly English words/hashtags and a few French ones, then our expectations is that were collecting mostly English and to a lesser extent French tweets. In further analysis, we restricted our work to English tweets.

For tweets pertaining to refugees, we filtered the aforementioned tweets using the following keyword: refugee, refuge, refugees, illegal, immigrant, immigrants, migrant, migrants, xenophobia, xenophobic, and xenophobe. The obtained subset of tweets was 137,113 tweets.  We sampled them to obtain a representative sample of 1,520 tweets that would result with a maximum +/- 2.5\% error rate at 95\% confidence. We tagged the tweets using 5 tags.  The tags were:\\ 
- pro-refugee: were the author expresses sympathy or support for refugees,\\ 
- anti-refugee: were the author attacks refugees, blames refugees for the terrorist acts, or calls from banning their entry,\\ 
- neutral: were the tweet is either not related or expresses mixed feelings,\\ 
- positive-news: were the tweet shares potentially pro-refugee news articles, and\\ 
- negative-news: were the tweet shares sharing potentially anti-refugee news articles.

The tagging was done by a single person who is well acquainted with politics in Western countries.  Table~\ref{annotRes} shows the breakdown of the five classes. As can be seen, the percentage of pro- and anti-refugee tweets is nearly identical.  However, the percentage of tweets containing references to news articles that could be damaging to the image of refugees was substantial.  Negative-news tweets were mostly dominated by stories claiming that at least one of the suicide bombers in the Paris attacks was a Syrian refugee who entered Europe through Greece.

Further analysis of prevalent hashtags appearing in pro- vs anti-refugee tweets can shed light on topics of interest for both groups.  To obtain better counts of hashtags, we automatically tagged all refugee related tweets from our initial set of 137,113 tweets that were nearly identical to any of the tagged tweets.  We matched tweets after removing name mentions, URLs, and lower-casing the tweets.  The number of tweets obtained after automatic tagging was 73,937.  On average, a pro-refugee user authored 1.13 tweets, while anti-refugee users authored 1.38 tweets.  We manually mapped all user supplied locations, which appeared for more than 3 tweets, to country names.   In all, we mapped the locations for 26,597 tweets to countries.  To determine which term hashtags are more prominent and more indicative of one class and not the other, We used the so-called ``Odds Ratio'' (OR).  Given:\\
A: the number of tweets in class 1 containing a term,\\ 
B: total number of tweets in class 1,\\ 
C: total number of tweets outside of class 1 containing term, and\\ 
D: total number of tweets outside of class 1, OR is computed as:\\
$OR = \frac{A * (D - C)}{C * (B - A)}$

\begin{table}
\begin{center}
\begin{tabular}{|p{2.5cm}|r|r|}
\hline
	& Count  &	Percentage \\\hline
pro-refugee	& 606	& 39.9\% \\
positive-news &	20	& 1.3\% \\
neutral	& 44	& 2.9\% \\
negative-news	& 163	& 10.7\% \\
anti-refugee	& 647	& 42.6\% \\\hline
Total	& 1520 & \\\hline
\end{tabular}
\end{center}
\caption{Breakdown of annotations per category.}
\label{annotRes}
\end{table}

The higher the value of OR, the more indicative/discriminating a term is for a particular class.  The top 25 hashtags for the tweets supporting refugees, ordered by OR value, where:
\#PorteOuverte (door open); \#BeirutAttacks; \#PrayForBeirut; \#PrayForPeace; \#Iraq; \#PrayForLebanon; \#G20; \#Terrorists; \#PeaceInParis; \#MuslimsAreNotTerrorist; \#TerroistAttack; \#Equality; \#Peace; \#SelectiveHumanity; \#IraqWar; \#Palestine; \#Poland; \#TodosSomosParis (we are all Paris); \#PrayForSyria; \#StayHuman; \#no2Rouhani; \#Syria; \#ISIL; \#AssadMustGo; \#ParrisAttacks; and \#UniteBlue.

As can be seen from the hashtags, there are: calls for an open door policy towards refugees (\#PorteOuverte and \#StayHuman); a mention of other places where terrorist attacks, mass killings, and wars have taken (or taking) place (ex. \#BeirutAttacks, \#PrayForBeirut, \#Iraq, and \#PrayForLebanon); a stance against \#ISIL (using acronym instead of ``Islamic State'' typically indicates antagonism against the group), Iran (\#no2Rouhani -- Rouhani is the Iranian president), the Assad regime in Syria (\#AssadMustGo); and to a lesser extent expression of US partisans politics (\#UniteBlue -- a left leaning organization).

The three most retweeted tweets were:\\
To people blaming refugees for attacks in Paris tonight. Do you not realise these are the people the refugees are trying to run away from..?\\
Over 200,000 people have died in Syria in the past 4.5 years.  That's a Paris attack EVERY SINGLE DAY.   That's what refugees are fleeing.\\
These murderers aren't refugees. Nor are they real Muslims.  They're terrorists who've hijacked a religion for nefarious gain. \#paris

These three tweets account for 46\% of the volume of pro-refugee tweets.  The most shared URLs were:
\begin{itemize}
\item a tweet by @jonsnowC4 that says: ``Paris police sources say the 2 Syrian passports found on the terrorists were fakes probably made in Turkey''\footnote{https://twitter.com/jonsnowC4/status/665814624672051201}
\item an AJ+ video featuring a Muslim saying that what ISIS does is not our Islam\footnote{https://amp.twimg.com/v/25797ac7-8e52-4993-b2f5-5b8d2925b8be}
\item a tweet by @EU\_Commission that says: ``Do not mix refugees with terrorists - @JunckerEU urges Europe not to give in to basic reactions. \#Paris \#G20''\footnote{https://twitter.com/EU\_Commission/status/665854381556764672}
\item a blog post by Aaron Y. Zelin entitled: ``The Islamic State on Refugees Leaving Syria''\footnote{http://jihadology.net/2015/11/14/the-islamic-state-on-refugees-leaving-syria/} which shows ISIS's antagonism to refugees fleeing Syria to the ``Abodes of the Unbelievers''.
\end{itemize}

The main topics discussed in the tweets were:\\
1. disassociating refugees from terrorism\\
2. blaming negative attitudes against refugees on racism and xenophobia\\
3. claiming that blaming refugees benefits ISIS\\
4. sharing news articles that discredit claims that one of the suicide bombers in the Paris attacks was a refugee\\
5. calling for prayers and unity

\begin{figure}[t]
\centering
\includegraphics[width=0.8 \columnwidth]{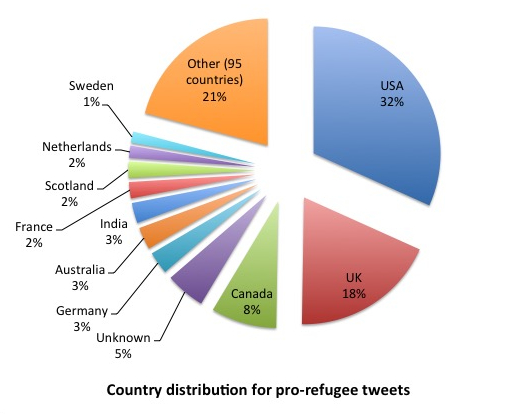}
\caption{\label{top-pro-refugee-loc}Geographical distribution of pro-refugee tweets}
\end{figure}

Figure\ref{top-pro-refugee-loc} shows the geographic distribution of pro-refugee tweets. As the figure shows, pro-refugee tweets come from 105 different countries, with 9 out of 10 of the top countries being Western countries.  It is noteworthy that tweets from Middle Eastern countries account for 4\% of the tweets.

The top 25 most indicative hashtags used in anti-refugee tweets were:
\#Greece; \#Migration; \#DemDebate; \#RefugeesNotWelcome; \#TCOT; \#NoRefugees; \#Anonymous; \#StopIslamicImmigration; \#Terrorist; \#SpecialReport;  \#ImpeachObama; \#ImpeachTheMuslim; \#BarrackHusseinObama;\#RedNationRising; \#Breaking; \#Germany; \#Muslim; \#Refugee; \#SecureTheBorder; \#RadicalIslam; \#LiberalLogic; \#Islam; \#Syrian; \#Election2016; \#IslamicState.

The hashtags followed along a few broad themes namely: the sharing of news articles that tie refugees to the Paris attacks, particular a story claiming that one of the suicide bombers entered Europe through Greece\footnote{The story was later discredited.} (\#Greece, \#SpecialReport, and \#Breaking); calls to shut out refugees (\#Migration, \#RefugeesNotWelcome, \#NoRefugees, and \#SecureTheBorder); US conservative partisanship (\#DemDebate (democratic debate), \#TCOT (top conservatives on Twitter), \#ImpeachObama,\#RedNationRising (a US conservative group), \#LiberalLogic, and \#Election2016); a strong anti-Muslim bias (ex. \#StopIslamicImmigration, \#RadicalIslam, and \#Islam) including claims that Barack Obama is a Muslim (ex. \#ImpeachTheMuslim and \#BarrackHusseinObama); and blame for \#Germany and other Western countries for opening their doors to refugees.

\begin{figure}[t]
\centering
\includegraphics[width=0.8 \columnwidth]{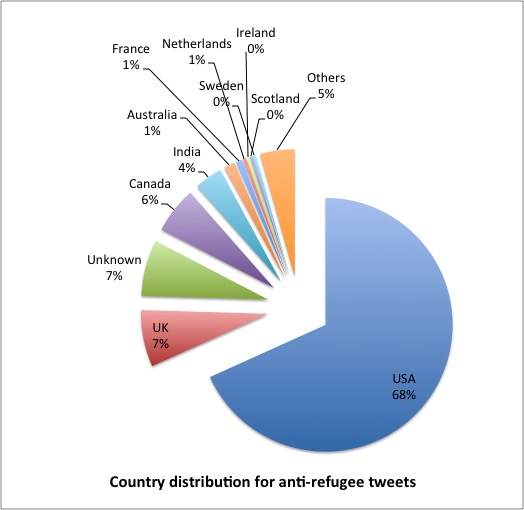}
\caption{\label{top-anti-refugee-loc}Geographical distribution of pro-refugee tweets}
\end{figure}

Figure\ref{top-anti-refugee-loc} shows the geographic distribution of anti-refugee tweets. As the figure shows, 68\% of the anti-refugee tweets come from the US, and the diversity of countries is smaller compared to the pro-refugee tweets (54 compared to 105 countries).  Tweets from Middle Eastern countries account for 0.3\% of the tweets with 73\% of them originating from Israel.

The three most tweeted anti-refugee tweets were:\\
We should not allow any political or religious group who want to destroy us and our way of life to immigrate to this country. \#ParisAttacks\\
Refugee \#migration issue. https://t.co/Yg5MR3Fw1p\\
We cannot allow Muslim immigrants to come across our borders unchecked while we are fighting this war on terror. \#ParisAttacks
 
These three tweets constitute 9.4\% of anti-refugee tweets.  The most shared URLs were:
\begin{itemize}
\item a tweet by @peterhook911 saying: ``How can France or Germany or any other euro union members be shocked.  How can these governments allow waves of unvetted young men in''\footnote{https://twitter.com/peterhook911/status/665661193416151040}
\item a Daily Mail article entitled: ``Revealed: Two of the Jihadis sneaked into Europe via Greece by posing as refugees and being rescued from a sinking migrant boat - and survivors say one of the attackers was a WOMAN''\footnote{http://www.dailymail.co.uk/news/article-3318379/Hunt-Isis-killers-Syrian-passport-body-suicide-bomber-Stade-France.html?ito=social-twitter\_mailonline}
\item a CNN article entitled: ``Belgium-born French national sought in Paris terror attacks''\footnote{http://edition.cnn.com/2015/11/15/world/paris-attacks/index.html?sr=twCNN111515paris-attacks1146PMVODtopLink\&linkId=18774297}
\item a Mirror article entitled: ``Paris attacks terrorist suspect Ahmed Almuhamed `was rescued near Greece after his refugee boat sunk'''\footnote{http://www.mirror.co.uk/news/world-news/paris-attacks-terrorist-suspect-ahmed-6836199}
\end{itemize}
From manually inspecting the tweets, the general themes used to attack refugees were the claims that:
\begin{itemize}
\item All refugees are young men
\item Refugees were not being vetted by authorities
\item ISIS fighters are disguised as refugees and one of the Paris attackers was a syrian refugee
\item Refugees are part of a Muslim invasion of Europe
\item Refugees are criminals who rape women and live off government subsidies
\item All Muslims are terrorist would spread terror in society
\item Left leaning politicians (ex. Obama, Clinton, Trudeau) continue to allow refugee as part of a larger plan to undermine Western societies.
\end{itemize}

A large percentage of anti-refugee tweets seem to express strict nationalistic/partisan views, where refugees and Muslims are considered as enemies. Specifically, they claim to support the republican party and oppose democratic politicians, where they claim that democratic politicians are ``opening the door to the enemy, who is carrying out a migration jihad''.

\section{Discussion and Conclusion}
In this paper we explored the immediate response of English speaking Twitter users towards refugees in the aftermath of the Paris terrorist attacks of Friday November 13, 2015.  The results show that percentage of pro-refugee and anti-refugee tweets are roughly equivalent.  However, there was a higher percentage of users sharing stories that could potentially harm attitudes towards the refugees.  The pro-refugee tweets seem to be significantly more geographically distributed than anti-refugee tweets.  Anti-refugee tweets were mostly coming from the US and were mostly partisan in nature and antagonistic to Muslims and refugees who were regarded as mostly Muslim.





\end{document}